\documentclass[conference]{IEEEtran}
\IEEEoverridecommandlockouts
\usepackage{cite}
\usepackage{amsmath,amssymb,amsfonts}
\usepackage{algorithm}
\usepackage{algpseudocode}
\usepackage{graphicx}
\usepackage{textcomp}
\usepackage{xcolor}

\def\BibTeX{{\rm B\kern-.05em{\sc i\kern-.025em b}\kern-.08em
    T\kern-.1667em\lower.7ex\hbox{E}\kern-.125emX}}
\begin{document}

\title{Complex-BP-Neural-Network-based Hybrid Precoding for Millimeter Wave Multiuser Massive MIMO Systems}

\author{\IEEEauthorblockN{Kai Chen\IEEEauthorrefmark{2}, Jing Yang\IEEEauthorrefmark{2}, Xiaohu Ge\IEEEauthorrefmark{2}\IEEEauthorrefmark{1}, Yonghui Li\IEEEauthorrefmark{3}}
\IEEEauthorblockA{\IEEEauthorrefmark{2}School of Electronic Information and Communications, Huazhong University of Science and Technology, Wuhan, Hubei,\\ China}
\IEEEauthorblockA{\IEEEauthorrefmark{3}School of Electrical and Information Engineering, University of Sydney, Sydney, Australia}
e-mail: xhge@mail.hust.edu.cn
}

\maketitle

\begin{abstract}
The high energy consumption of massive multi-input multi-out (MIMO) system has become a prominent problem in the millimeter wave(mm-Wave) communication scenario. The hybrid precoding technology greatly reduces the number of radio frequency(RF) chains by handing over part of the coding work to the phase shifting network, which can effectively improve energy efficiency. However, conventional hybrid precoding algorithms based on mathematical means often suffer from performance loss and high computational complexity. In this paper, a novel BP-neural-network-enabled hybrid precoding algorithm is proposed, in which the full-digital zero-forcing(ZF) precoding is set as the training target. Considering that signals at the base station are complex, we choose the complex neural network that has a richer representational capacity. Besides, we present the activation function of the complex neural network and the gradient derivation of the back propagation process. Simulation results demonstrate that the performance of the proposed hybrid precoding algorithm can optimally approximate the ZF precoding.
\end{abstract}


\section{Introduction}
The explosive growth of data have brought enormous challenges to the current communication field. Advanced signal transceiver technologies are required to provide higher capacity \cite{b22,b23,b25}. Massive multi-input multi-out(MIMO) is a technology with a large number of antennas at the base station, which can provide extremely high beamforming gain to compensate for the high path loss of the millimeter wave(mm-Wave) channel \cite{b2}. Combined with the mm-Wave featuring large bandwidth, massive MIMO have become a promising technology for the future mm-Wave communication systems.

The massive MIMO systems can increase communication performance significantly, but the large number of radio frequency(RF) chains at the base station will consume huge energy. Reduced-RF based hybrid precoding is one of the effective solutions to this problem \cite{b26,b27,b28}. Hybrid precoding consists of baseband precoding and RF precoding. The RF precoding is implemented by a phase shifting network where only the phase of the transmitting signals can be adjusted. The baseband precoding is performed at the baseband stage, where both the amplitude and the phase of the signals are adjustable \cite{b3}. There have been rich studies on hybrid precoding. In \cite{b4}, a hierarchical multi-resolution codebook was developed, and then a codebook-based hybrid precoding scheme was designed based on the channel estimation results. Authors in \cite{b5} presented a low-complexity precoding scheme named phased-ZF(PZF) precoding, where the baseband ZF precoding was determined by the equivalent channel seen from the baseband. Hybrid precoding based on SVD decomposition in single-user and multiuser scenarios was proposed in \cite{b6} and \cite{b7} respectively. In \cite{b8}, the analog precoder design problem for the frequency selective channel is transformed into a narrowband analog precoder design problem, and the baseband precoder is decomposed into two matrices: one is used to approach the optimal precoders jointly with the RF precoder, the other is used to eliminate multiuser interference via the ZF method. Considering the sparse scattering characteristics of millimeter-wave channels, spatial sparse precoding based on orthogonal matching pursuit (OMP) algorithm has also been studied extensively \cite{b9}.

Although the aforementioned hybrid precoding algorithms based on traditional mathematical models can achieve good performance, its high computational complexity will bring high energy consumption and delay \cite{b12}. With the rapid development of artificial intelligence (AI) technology, the neural network principle provides the possibility of surpassing traditional methods for the design and optimization of 5G systems benefiting from its powerful information extraction capabilities. There have been some works combining artificial intelligence with mobile communication systems, like RF resource allocation, non-orthogonal multiple access (NOMA), optimal reception and channel estimation \cite{b14,b15,b16,b17,b19}. However, few researches have been done to solve the hybrid precoding problem via AI technology. In \cite{b20}, a deep-learning-enabled hybrid precoding model for the single-user massive MIMO system was proposed for the first time. The deep neural network framework included the transceiver end and the channel of the communication system, but only the hybrid precoding at the transmitting end was trained in the paper.

In view of the powerful ability of neural network to process large amounts of data and to solve non-convex optimization problems, this paper proposes a hybrid precoding algorithm based on BP neural network in the mm-Wave multiuser massive MIMO system. Normally, transmitting signals at the base station belong to the complex domain, so a complex neural network is adopted, while \cite{b20} uses the real neural network. The split activation function is chosen for nonlinear mapping and the power limitation of precoding. The detailed gradient derivation of the back propagation process of the neural network is also given. Simulation results show that the proposed hybrid precoding algorithm based on complex neural network is better than the traditional algorithms based on mathematical models, and can optimally approximate full digital precoding. To the best of our knowledge, this paper is the first to incorporate a complex neural network to hybrid precoding.

The remainder of the article is organized as follows. The system model is introduced in Section II. In Section III, the hybrid precoding scheme based on complex BP neural network is proposed, and the corresponding network model and algorithm are given. Then, the performance of the proposed algorithm is analyzed in Section IV. Finally, we summarize the paper in Section V.

\emph{Notations:} Bold uppercase $\textbf{M}$ is a matrix, bold lowercase $\textbf{m}$ is a vector, and $\Omega$ is a set. $\left[\bullet\right]^T$ and $\left[\bullet\right]^H$ denote transpose and conjugate transpose respectively. $\left\|\bullet\right\|_F$ represents the Frobenius norm of the matrix. $z_n^{\left(l\right)}$ is the activation input of the $n$th neuron of the $\left({l + 1}\right)$th layer. $a_m^{\left(l\right)}$ is the activation output of the $m$th neuron of the $l$th layer. $w_{nm}^{\left(l\right)}$ is the synaptic weight of the $n$th neuron of the $\left({l + 1}\right)$th layer relative to the  $m$th output of the $l$th layer.
\section{System Model}
\begin{figure}[htbp]
\centerline{\includegraphics[width=0.5\textwidth]{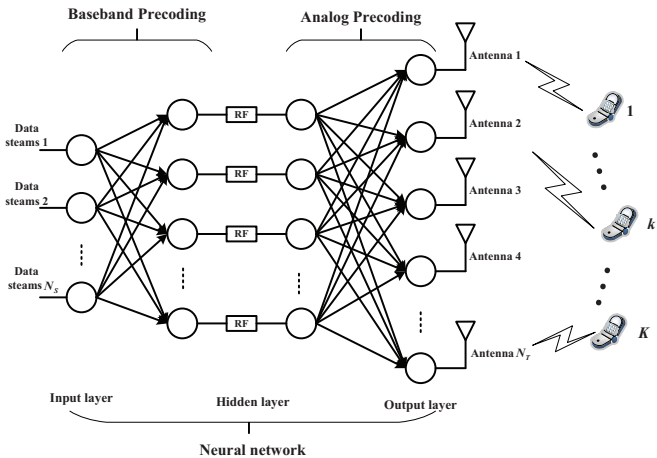}}
\caption{System model of the proposed hybrid precoding scheme.}
\label{fig1}
\end{figure}

In this paper, we consider a typical mm-Wave multiuser massive MIMO communication system as shown in Fig.~\ref{fig1}. $N_S$ data streams is transmitted via $N_T$ antennas at the base station, where the $N_{RF}$ RF chains satisfy ${N_S}\le{N_{RF}}<{N_T}$. The number of user equipments(UEs) at the receiving end is $K=N_S$.

The received signal $y_k$ of the $k$th user is

\begin{equation}
{y_k} = \textbf{h}_k^H\textbf{x}+{n_k},
\label{eq1}
\end{equation}
where ${n_k} \sim {\cal C}{\cal N}\left( {0,\sigma _n^2} \right)$ represents the additive white Gaussian noise (AWGN) and $\sigma _n^2$ is the noise power. $\textbf{x} \in {{\mathbb{C}}^{{{N}_{\text{T}}}\times1}}$ is the transmitting signals of the antennas, which is generated by the source amplitude-phase domain data streams $\textbf{s} = {\left[ {{s_1},{s_2},...,{s_{{N_S}}}} \right]^H}$ via hybrid precoding, and $\mathbb{E}\left[ {\textbf{s}{\textbf{s}^H}} \right] = 1$ . The precoding process can be expressed as

\begin{equation}
\textbf{x} = \textbf{ADs} = \textbf{Fs},
\label{eq2}
\end{equation}
where $\textbf{D} \in {\mathbb{C}^{{N_{RF}} \times {N_S}}}$ is the baseband precoding matrix and $\textbf{A} \in {\mathbb{C}^{{N_T} \times {N_{RF}}}}$ is the RF precoding matrix. In this paper, we use BP neural network $\textbf{F} \in {\mathbb{C}^{{N_T} \times {N_S}}}$ for hybrid precoding. Noticeably, $\textbf{F}$ is not a real matrix, but the coding process of the neural network, which can be seen as the target full-digital precoding matrix. Due to the limitation of the transmitting power ${P_{\max }}$ at the base station, the hybrid precoding matrix should satisfy

\begin{equation}
\left\| \textbf{F} \right\|_F^2 = \left\| \textbf{AD} \right\|_F^2 \leqslant {P_{\max }}.
\label{eq3}
\end{equation}

The downlink mm-Wave channel matrix is ${\textbf{H}^H} = {\left[ {\textbf{h}_1,...,\textbf{h}_k,...,\textbf{h}_K} \right]^H} \in {\mathbb{C}^{K \times {N_T}}}$, where $\textbf{h}_k \in {\mathbb{C}^{{N_T} \times 1}}$ is the channel vector from the base station to the $k$th user. Considering the sparse property of the mm-Wave channel and the high correlation of large antenna arrays, we adopt a geometric channel model with limited scattering and multipath \cite{b3,b9}. Then each $\textbf{h}_k$ can be expressed as

\begin{equation}
{{\textbf{h}}_k} = \sqrt {\frac{{{N_T}{\beta _k}}}{{{N_{ray}}}}} \sum\limits_{i = 1}^{{N_{ray}}} {{\rho _{k,i}}\textbf{u}\left( {{\psi _{k,i}},{\vartheta _{k,i}}} \right)} ,
\label{eq4}
\end{equation}
where $N_{ray}$ is the number of the multipath. ${\beta _k} = \zeta /l_k^\gamma $ denotes the large scale fading coefficient between the base station and the $k$th user, where $\zeta $ obeys a log-normal distribution with a 0-mean and 9.2dB-variance. ${l_k}$ is the distance between the base station and the $k$th user and $\gamma $ is the path loss factor. ${\rho _{k,i}} \sim \mathcal{C}\mathcal{N}\left( {0,\sigma _{k,i}^2} \right)$ is the complex gain on the $i$th multipath. ${\psi _{k,i}}$ and ${\vartheta _{k,i}}$ represent the azimuth and elevation angle of departure respectively. The uniform planar array(UPA) is employed in this paper and the array response $\textbf{u}\left( {{\psi _{k,i}},{\vartheta _{k,i}}} \right)$ is

\begin{equation}
\begin{gathered}
  {\textbf{u}}\left( {{\psi _{k,i}},{\vartheta _{k,i}}} \right) = \frac{1}{{\sqrt {{N_T}} }}\left[ {1,.} \right...,\exp j\frac{{2\pi }}{\lambda }d\left( {l\sin \left( {{\psi _{k,i}}} \right)sin\left( {{\vartheta _{k,i}}} \right)} \right. \hfill \\
  \;\;\;\;\;\;\left. { + rcos\left( {{\vartheta _{k,i}}} \right)} \right),...,\exp j\frac{{2\pi }}{\lambda }d\left( {\left( {L - 1} \right)\sin \left( {{\psi _{k,i}}} \right)sin\left( {{\vartheta _{k,i}}} \right)} \right. \hfill \\
  \;\;\;\;\;\;{\left. { + \left. {\left( {R - 1} \right)cos\left( {{\vartheta _{k,i}}} \right)} \right)} \right]^T}, \hfill \\
\end{gathered}
\label{eq5}
\end{equation}
where $\lambda $ represents the wavelength of the mm-Wave and $d = \lambda /2$ is the inter-element spacing. $0 \leqslant l \leqslant \left( {L - 1} \right)$ and $0 \leqslant r \leqslant \left( {R - 1} \right)$ are the row and column indices of the antenna array respectively and the antenna array size is ${N_T} = LR$.

\section{Proposed Scheme}
It is obvious that the hybrid precoding of the massive MIMO system shares a similar topology with the neural network. The signal processing of the hybrid precoding can be seen as the matrix multiplication operation, which is also analogous to the weight processing in a neural network. In addition, most of the previous hybrid precoding algorithms are aimed at approximating the full digital precoding, which is similar to the way of training a neural network via reducing the cost function. Therefore, we consider using the BP neural network to design the hybrid precoding. This section will give the detailed model of the neural network architecture.

\subsection{Neural Network Architecture}\label{AA}
Numerous studies have shown that complex neural networks have better performance in signal processing because the phase of complex signals encodes shape, edge and direction. At the same time, complex weights in neural networks is biologically significant. The firing rate and the relative timing of human brain activity are correspond to the amplitude and phase of complex neurons respectively \cite{b21}. Therefore, a complex BP neural network is adopted to perform hybrid precoding on signals.

\begin{figure}[htbp]
\centerline{\includegraphics[width=0.5\textwidth]{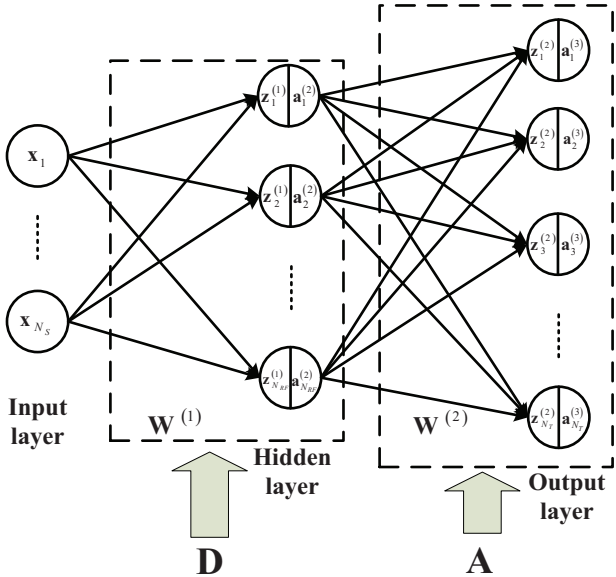}}
\caption{Architecture of neural network for hybrid precoding.}
\label{fig2}
\end{figure}

The architecture of the neural network is depicted in Fig.~\ref{fig2}. In order to correspond to the physical structure of the hybrid precoding, a single hidden layer is used. The input layer contains ${N_S}$ neurons, corresponding to the input data streams. The hidden layer contains ${N_{RF}}$ neurons, corresponding to the RF chains. The output layer contains ${N_{T}}$ neurons, corresponding to the transmitting antennas. Consequently, the weight matrix ${\textbf{W}^{\left( 1 \right)}}$ and ${\textbf{W}^{\left( 2 \right)}}$ along with the activation function play the role of the baseband precoding and the RF precoding, respectively.

With the optimal full-digital ZF precoding as the training target, we construct a training set. The input of the training sample is the ${N_S} \times 1$ dimension source signal streams, and the output is the signal streams ${\textbf{y}_{ZF}} \in {\mathbb{C}^{{N_T} \times 1}}$ to be transmitted after ZF precoding. The cost function of the neural network can be expressed as

\begin{equation}
{e^2} = \frac{1}{2}\left\| {{\textbf{a}^{\left( 3 \right)}} - {\textbf{y}_{ZF}}} \right\|_F^2.
\label{eq6}
\end{equation}

In our neural network, all the signals are complex during iteration process, which causes the activation function to be a complex one. Besides, the activation function should have the ability of power limitation in addition to nonlinear mapping. This requires the activation function to satisfy both the derivable and bounded conditions. According to the Cauchy-Riemann equations and the Liouville¡¯s theorem, we adopt the split activation function:

\begin{equation}
f\left( z \right) = r\left( {isgm\left( x \right) + j \cdot isgm\left( y \right)} \right),
\label{eq7}
\end{equation}
where $r = \sqrt {{P_{\max }}/2} $ is the power limitation factor. $isgm\left( x \right) \in \left( { - 1,1} \right)$ is the improved sigmod function:

\begin{equation}
isgm\left( x \right) = \frac{2}{{1 + {e^{ - x}}}} - 1.
\label{eq8}
\end{equation}

The derivative of $isgm\left( x \right)$ is $isgm'\left( x \right) = \left[ {1 + isgm\left( x \right)} \right]\left[ {1 - isgm\left( x \right)} \right]/2$.
\subsection{Training Process}
The training process of BP neural network can be divided into two parts: forward propagation and back propagation. The forward propagation is summarized as

\begin{equation}
z_n^{\left( 1 \right)} = \sum\limits_{m = 1}^{{N_S}} {w_{nm}^{\left( 1 \right)}{x_m}}
\label{eq9}
\end{equation}

\begin{equation}
a_n^{\left( 2 \right)} = f\left( {z_n^{\left( 1 \right)}} \right)
\label{eq10}
\end{equation}

\begin{equation}
z_n^{\left( 2 \right)} = \sum\limits_{m = 1}^{{N_{RF}}} {w_{nm}^{\left( 2 \right)}a_m^{\left( 2 \right)}}
\label{eq11}
\end{equation}

\begin{equation}
a_n^{\left( 3 \right)} = f\left( {z_n^{\left( 2 \right)}} \right).
\label{eq12}
\end{equation}

The improved momentum algorithm of stochastic gradient descent(SGD) is adopted to update the weights during the back propagation process:

\begin{equation}
w_{nm}^{\left( l \right)}\left( {k + 1} \right) = w_{nm}^{\left( l \right)}\left( k \right) - \Delta w_{nm}^{\left( l \right)}\left( k \right)
\label{eq13}
\end{equation}

\begin{equation}
\Delta w_{nm}^{\left( l \right)}\left( k \right) = \alpha \Delta w_{nm}^{\left( l \right)}\left( {k - 1} \right) + \mu {\nabla _{w_{nm}^{\left( l \right)}}}\left( {{e^2}} \right),
\label{eq14}
\end{equation}
where $\alpha $ is the momentum factor and $\mu $ is the learning rate.

The BP algorithm adjusts the weight by calculating the gradient ${\nabla _{w_{nm}^{\left( l \right)}}}\left( {{e^2}} \right)$ of the cost function with respect to the weight. Since the cost function ${e^2}$ is a real function and is not analytic, we need to solve the partial derivatives of the cost function with respect to both the real and imaginary parts of the weight $w_{nm}^{\left( l \right)}$ separately. Weight $w_{nm}^{\left( l \right)}$ is written as

\begin{equation}
w_{nm}^{\left( l \right)} = w_{Rnm}^{\left( l \right)} + j \cdot w_{Inm}^{\left( l \right)}.
\label{eq15}
\end{equation}

And the gradient is

\begin{equation}
{\nabla _{w_{nm}^{\left( l \right)}}}\left( {{e^2}} \right) = \frac{{\partial {e^2}}}{{\partial w_{Rnm}^{\left( l \right)}}} + j \cdot \frac{{\partial {e^2}}}{{\partial w_{Inm}^{\left( l \right)}}}.
\label{eq16}
\end{equation}

Using the chain rule, the gradient of weights at the second layer is

\begin{equation}
\begin{gathered}
  \frac{{\partial {e^2}}}{{\partial w_{Rnm}^{\left( 2 \right)}}} = \frac{{\partial {e^2}}}{{\partial a_{Rn}^{\left( 3 \right)}}}\frac{{\partial a_{Rn}^{\left( 3 \right)}}}{{\partial z_{Rn}^{\left( 2 \right)}}}\frac{{\partial z_{Rn}^{\left( 2 \right)}}}{{\partial w_{Rnm}^{\left( 2 \right)}}} + \frac{{\partial {e^2}}}{{\partial a_{In}^{\left( 3 \right)}}}\frac{{\partial a_{In}^{\left( 3 \right)}}}{{\partial z_{In}^{\left( 2 \right)}}}\frac{{\partial z_{In}^{\left( 2 \right)}}}{{\partial w_{Rnm}^{\left( 2 \right)}}} \hfill \\
  \;\;\;\;\;\;\;\;\;\;\;\; = r\left[ {\left( {a_{Rn}^{\left( 3 \right)} - {y_{Rn}}} \right)isgm'\left( {z_{Rn}^{\left( 2 \right)}} \right)a_{Rm}^{\left( 2 \right)}}\right. \hfill \\
  \;\;\;\;\;\;\;\;\;\;\;\;\;\;\;\left.{ +\left( {a_{In}^{\left( 3 \right)} - {y_{In}}} \right)isgm'\left( {z_{In}^{\left( 2 \right)}} \right)a_{Im}^{\left( 2 \right)}} \right] \hfill \\
\end{gathered}
\label{eq17}
\end{equation}

\begin{equation}
\begin{gathered}
  \frac{{\partial {e^2}}}{{\partial w_{Inm}^{\left( 2 \right)}}} = \frac{{\partial {e^2}}}{{\partial a_{Rn}^{\left( 3 \right)}}}\frac{{\partial a_{Rn}^{\left( 3 \right)}}}{{\partial z_{Rn}^{\left( 2 \right)}}}\frac{{\partial z_{Rn}^{\left( 2 \right)}}}{{\partial w_{Inm}^{\left( 2 \right)}}} + \frac{{\partial {e^2}}}{{\partial a_{In}^{\left( 3 \right)}}}\frac{{\partial a_{In}^{\left( 3 \right)}}}{{\partial z_{In}^{\left( 2 \right)}}}\frac{{\partial z_{In}^{\left( 2 \right)}}}{{\partial w_{Inm}^{\left( 2 \right)}}} \hfill \\
  \;\;\;\;\;\;\;\;\;\;\;\;= r\left[ { - \left( {a_{Rn}^{\left( 3 \right)} - {y_{Rn}}} \right)isgm'\left( {z_{Rn}^{\left( 2 \right)}} \right)a_{Im}^{\left( 2 \right)}}\right. \hfill \\
  \;\;\;\;\;\;\;\;\;\;\;\;\;\;\;\left.{+ \left( {a_{In}^{\left( 3 \right)} - {y_{In}}} \right)isgm'\left( {z_{In}^{\left( 2 \right)}} \right)a_{Rm}^{\left( 2 \right)}} \right], \hfill \\
\end{gathered}
\label{eq18}
\end{equation}
where ${y_i}$ is the $i$th element of the output of training sample ${\textbf{y}_{ZF}}$. Similarly, the gradient of weights at the first layer is

\begin{equation}
\begin{gathered}
  \frac{{\partial {e^2}}}{{\partial w_{Rnm}^{\left( 1 \right)}}} = \sum\limits_{i = 1}^{{N_T}} {\left[ {\frac{{\partial {e^2}}}{{\partial a_{Ri}^{\left( 3 \right)}}}\frac{{\partial a_{Ri}^{\left( 3 \right)}}}{{\partial z_{Ri}^{\left( 3 \right)}}}\left( {\frac{{\partial z_{Ri}^{\left( 3 \right)}}}{{\partial a_{Rn}^{\left( 2 \right)}}}\frac{{\partial a_{Rn}^{\left( 2 \right)}}}{{\partial z_{Rn}^{\left( 2 \right)}}}\frac{{\partial z_{Rn}^{\left( 2 \right)}}}{{\partial w_{Rnm}^{\left( 2 \right)}}}} \right.} \right.}  \hfill \\
  \;\;\;\;\;\;\;\;\;\;\;\;\;\;\left. { + \frac{{\partial z_{Ri}^{\left( 3 \right)}}}{{\partial a_{In}^{\left( 2 \right)}}}\frac{{\partial a_{In}^{\left( 2 \right)}}}{{\partial z_{In}^{\left( 2 \right)}}}\frac{{\partial z_{In}^{\left( 2 \right)}}}{{\partial w_{Rnm}^{\left( 2 \right)}}}} \right) \hfill \\
  \;\;\;\;\;\;\;\;\;\;\;\;\;\;+ \frac{{\partial {e^2}}}{{\partial a_{Ii}^{\left( 3 \right)}}}\frac{{\partial a_{Ii}^{\left( 3 \right)}}}{{\partial z_{Ii}^{\left( 3 \right)}}}\left( {\frac{{\partial z_{Ii}^{\left( 3 \right)}}}{{\partial a_{Rn}^{\left( 2 \right)}}}\frac{{\partial a_{Rn}^{\left( 2 \right)}}}{{\partial z_{Rn}^{\left( 2 \right)}}}\frac{{\partial z_{Rn}^{\left( 2 \right)}}}{{\partial w_{Rnm}^{\left( 2 \right)}}}} \right. \hfill \\
  \;\;\;\;\;\;\;\;\;\;\;\;\;\;\left. {\left. { + \frac{{\partial z_{Ii}^{\left( 3 \right)}}}{{\partial a_{In}^{\left( 2 \right)}}}\frac{{\partial a_{In}^{\left( 2 \right)}}}{{\partial z_{In}^{\left( 2 \right)}}}\frac{{\partial z_{In}^{\left( 2 \right)}}}{{\partial w_{Rnm}^{\left( 2 \right)}}}} \right)} \right] \hfill \\
   = {r^2}\sum\limits_{i = 1}^{{N_T}} {\left[ {\left( {a_{Ri}^{\left( 3 \right)} - {y_{Ri}}} \right)isgm'\left( {z_{Ri}^{\left( 3 \right)}} \right)\left( {w_{Rin}^{\left( 3 \right)}isgm'\left( {z_{Rn}^{\left( 2 \right)}} \right){x_{Rm}}} \right.} \right.}  \hfill \\
  \;\;\;\;\;\;\;\;\;\;\;\;\;\;\left. { - w_{Iin}^{\left( 3 \right)}isgm'\left( {z_{In}^{\left( 2 \right)}} \right){x_{Im}}} \right) \hfill \\
  \;\;\;\;\;\;\;\;\;\;\;\;\;\;+ \left( {a_{Ii}^{\left( 3 \right)} - {y_{Ii}}} \right)isgm'\left( {z_{Ii}^{\left( 3 \right)}} \right)\left( {w_{Iin}^{\left( 3 \right)}isgm'\left( {z_{Rn}^{\left( 2 \right)}} \right){x_{Rm}}} \right. \hfill \\
  \;\;\;\;\;\;\;\;\;\;\;\;\;\;\left. {\left. { + w_{Rin}^{\left( 3 \right)}isgm'\left( {z_{In}^{\left( 2 \right)}} \right){x_{Im}}} \right)} \right] \hfill \\
\end{gathered}
\label{eq19}
\end{equation}

\begin{equation}
\begin{gathered}
\frac{{\partial {e^2}}}{{\partial w_{Inm}^{\left( 1 \right)}}} = \sum\limits_{i = 1}^{{N_T}} {\left[ {\frac{{\partial {e^2}}}{{\partial a_{Ri}^{\left( 3 \right)}}}\frac{{\partial a_{Ri}^{\left( 3 \right)}}}{{\partial z_{Ri}^{\left( 3 \right)}}}\left( {\frac{{\partial z_{Ri}^{\left( 3 \right)}}}{{\partial a_{Rn}^{\left( 2 \right)}}}\frac{{\partial a_{Rn}^{\left( 2 \right)}}}{{\partial z_{Rn}^{\left( 2 \right)}}}\frac{{\partial z_{Rn}^{\left( 2 \right)}}}{{\partial w_{Inm}^{\left( 2 \right)}}}} \right.} \right.}  \hfill \\
  \;\;\;\;\;\;\;\;\;\;\;\;\;\;\left. { + \frac{{\partial z_{Ri}^{\left( 3 \right)}}}{{\partial a_{In}^{\left( 2 \right)}}}\frac{{\partial a_{In}^{\left( 2 \right)}}}{{\partial z_{In}^{\left( 2 \right)}}}\frac{{\partial z_{In}^{\left( 2 \right)}}}{{\partial w_{Inm}^{\left( 2 \right)}}}} \right) \hfill \\
  \;\;\;\;\;\;\;\;\;\;\;\;\;\;+ \frac{{\partial {e^2}}}{{\partial a_{Ii}^{\left( 3 \right)}}}\frac{{\partial a_{Ii}^{\left( 3 \right)}}}{{\partial z_{Ii}^{\left( 3 \right)}}}\left( {\frac{{\partial z_{Ii}^{\left( 3 \right)}}}{{\partial a_{Rn}^{\left( 2 \right)}}}\frac{{\partial a_{Rn}^{\left( 2 \right)}}}{{\partial z_{Rn}^{\left( 2 \right)}}}} \right.\frac{{\partial z_{Rn}^{\left( 2 \right)}}}{{\partial w_{Inm}^{\left( 2 \right)}}} \hfill \\
  \;\;\;\;\;\;\;\;\;\;\;\;\;\;\left. {\left. { + \frac{{\partial z_{Ii}^{\left( 3 \right)}}}{{\partial a_{In}^{\left( 2 \right)}}}\frac{{\partial a_{In}^{\left( 2 \right)}}}{{\partial z_{In}^{\left( 2 \right)}}}\frac{{\partial z_{In}^{\left( 2 \right)}}}{{\partial w_{Inm}^{\left( 2 \right)}}}} \right)} \right] \hfill \\
   = {r^2}\sum\limits_{i = 1}^{{N_T}} {\left[ { - \left( {a_{Ri}^{\left( 3 \right)} - {y_{Ri}}} \right)isgm'\left( {z_{Ri}^{\left( 3 \right)}} \right)\left( {w_{Rin}^{\left( 3 \right)}isgm'\left( {z_{Rn}^{\left( 2 \right)}} \right){x_{Im}}} \right.} \right.}  \hfill \\
  \;\;\;\;\;\;\;\;\;\;\;\;\;\;\left. { + w_{Iin}^{\left( 3 \right)}isgm'\left( {z_{In}^{\left( 2 \right)}} \right){x_{Rm}}} \right) \hfill \\
  \;\;\;\;\;\;\;\;\;\;\;\;\;\;+ \left( {a_{Ii}^{\left( 3 \right)} - {y_{Ii}}} \right)isgm'\left( {z_{Ii}^{\left( 3 \right)}} \right)\left( { - w_{Iin}^{\left( 3 \right)}isgm'\left( {z_{Rn}^{\left( 2 \right)}} \right){x_{Im}}} \right. \hfill \\
  \;\;\;\;\;\;\;\;\;\;\;\;\;\;\left. {\left. { + w_{Rin}^{\left( 3 \right)}isgm'\left( {z_{In}^{\left( 2 \right)}} \right){x_{Rm}}} \right)} \right] \hfill \\
\end{gathered}
\label{eq20}
\end{equation}

\subsection{Algorithm Summary}
The training process of the proposed hybrid precoding network is similar to that of the traditional BP neural network. First, the complex weight is initialized. Then forward propagation is performed to obtain the corresponding network output and the cost function. Afterwards, the partial derivatives are calculated according to \eqref{eq17},\eqref{eq18},\eqref{eq19} and \eqref{eq20} to adjust the weight. The training process is terminated when the error

\begin{algorithm}[htbp]
\caption{Hybrid precoding algorithm based on BP neural network.}
\begin{algorithmic}[1]

\Require  Channel matrix $\textbf{H}$
\Ensure Optimized hybrid precoding neural network $\textbf{F}$
\State Initialization: Weights of the neural network are initialized as ${\textbf{W}^{\left( 1 \right)}},{\textbf{W}^{\left( 2 \right)}} \sim \mathcal{C}\mathcal{N}\left( {0,1} \right)$ and epoch is 1. The threshold of the error is set as ${10^{ - 8}}$;
\State Calculate the optimal full-digital ZF precoding matrix ${\textbf{B}_{ZF}}$ from the channel matrix;
\State Generate the training set ${\Omega _t}\left( {{\textbf{x}_i},{\textbf{y}_i}} \right)$ and the test set ${\Omega _v}\left( {{\textbf{x}_i},{\textbf{y}_i}} \right)$. The input of both set is ${\textbf{x}_i} \sim \mathcal{C}\mathcal{N}\left( {0,1} \right)$ and the output is ${\textbf{y}_i} = {\textbf{B}_{ZF}}{\textbf{x}_i}$;
\While{epoch$\leqslant$200}
\State Train the neural network: Perform the forward propagation according to \eqref{eq9},\eqref{eq10},\eqref{eq11} and \eqref{eq12} and calculate the cost function ${e^2}$;
\State Calculate the gradient ${\nabla _{w_{nm}^{\left( l \right)}}}\left( {{e^2}} \right)$ according to \eqref{eq16},\eqref{eq17},\eqref{eq18},\eqref{eq19} and \eqref{eq20};
\State Perform the back propagation via SGD and update the weights according to \eqref{eq12},\eqref{eq13} and \eqref{eq14};
\State Calculate the error of the test set. If the error is smaller than the threshold, skip to step 10;
\EndWhile
\State \Return Optimized hybrid precoding neural network $\textbf{F}$.
\end{algorithmic}
\end{algorithm}

\noindent falls to an acceptable range. The algorithm is summarized in Algorithm 1.

For a certain channel, the training set of the network is generated autonomously. Therefore, the network supports online learning. And the weights are adjusted in real time based on changes of the channel.

\section{Simulation Results}
In this section, we present the simulation results of the hybrid precoding scheme based on BP neural network. We compare the performance of our proposed algorithm with full-digital ZF precoding, PZF precoding \cite{b5}, and spatially sparse precoding based on OMP algorithm \cite{b9} on spectrum efficiency and bit error rate(BER). Without special instructions, ${N_T} = 128$, ${N_{RF}} = 16$, ${N_{ray}} = 80$, the number of samples in the training set is 100, and the maximum of training iterations is 200.

\subsection{Spectrum Efficiency}
The spectrum efficiency of a mm-Wave multiuser massive MIMO system is \cite{b24}

\begin{equation}
R = \sum\limits_{k = 1}^K {{{\log }_2}\left( {1 + SIN{R_k}} \right)},
\label{eq21}
\end{equation}
where $SIN{R_k}$ is the signal to interference and noise ratio(SINR) of the  $k$th user at the receiving end, which can be expressed as

\begin{equation}
SIN{R_k} = \frac{{\textbf{h}_k^H\textbf{A}{\textbf{d}_k}\textbf{d}_k^H{\textbf{A}^H}{\textbf{h}_k}}}{{\sigma _n^2 + \sum\limits_{i = 1,i \ne k}^K {\textbf{h}_k^H\textbf{A}{\textbf{d}_i}\textbf{d}_i^\textbf{H}{\textbf{A}^H}{\textbf{h}_k}} }},
\label{eq22}
\end{equation}
where ${\textbf{d}_j}$ is the  $j$th column of $\textbf{D}$.

When it comes to our proposed hybrid precoding algorithm based on BP neural network, we calculate the spectrum efficiency by using the unit matrix ${\textbf{I}_K}$ as the input of the neural network. Then the output of the neural network is $\textbf{Y} = \left[ {{\textbf{y}_1},...,{\textbf{y}_k},...,{\textbf{y}_K}} \right] \in {C^{{N_T} \times K}}$. Furthermore, the SINR of the  $k$th user can be expressed as

\begin{equation}
SIN{R_k} = \frac{{\textbf{h}_k^H{\textbf{y}_k}\textbf{y}_k^H{\textbf{h}_k}}}{{\sigma _n^2 + \sum\limits_{i = 1,i \ne k}^K {\textbf{h}_k^H{\textbf{y}_i}\textbf{y}_i^H{\textbf{h}_k}} }}.
\label{eq23}
\end{equation}

\subsection{Performance Analysis}
\begin{figure}[htbp]
\centerline{\includegraphics[width=0.5\textwidth]{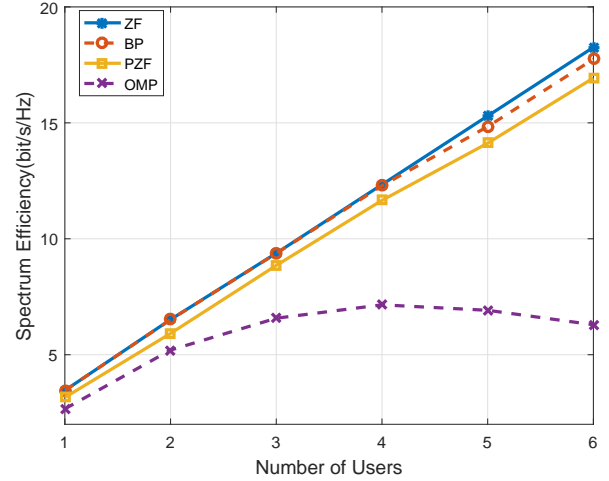}}
\caption{Spectrum efficiency versus the number of users.}
\label{fig3}
\end{figure}

Fig.~\ref{fig3} shows the spectrum efficiency achieved by the proposed hybrid precoding algorithm as well as ZF precoding, PZF precoding, and OMP-based precoding versus the number of users. Firstly, it¡¯s apparent that our proposed algorithm outperforms the OMP and PZF algorithm, and can better approximate the spectrum efficiency of full digital precoding, which benefits from the powerful information extraction and representation capabilities of the neural network. Secondly, as the number of users increases, the performance gap between the hybrid precoding algorithms and the full digital precoding becomes larger, but our proposed algorithm has a smaller

\begin{figure}[htbp]
\centerline{\includegraphics[width=0.5\textwidth]{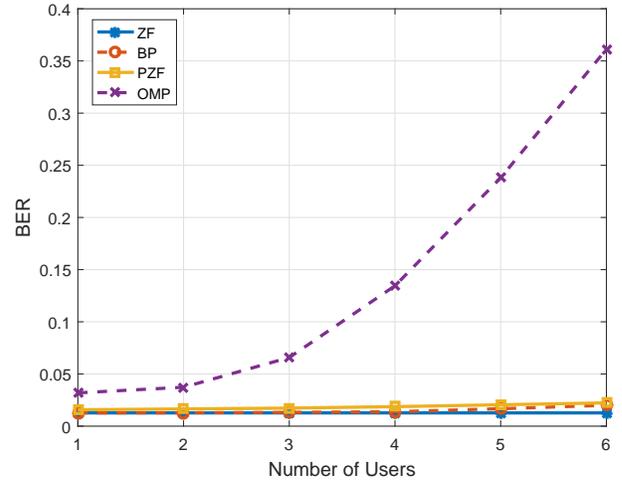}}
\caption{BER versus the number of users.}
\label{fig4}
\end{figure}

\noindent performance loss. This is because more users cause more serious inter-user interference, and precoding algorithms based on mathematical methods can hardly eliminate this negative impact completely. For a neural network, more users mean an increase in neuron nodes and weights, so more features are needed to be extracted to construct a better network. In fact, the performance degradation of our proposed algorithm can be improved by adding the training samples.

\begin{figure}[htbp]
\centerline{\includegraphics[width=0.5\textwidth]{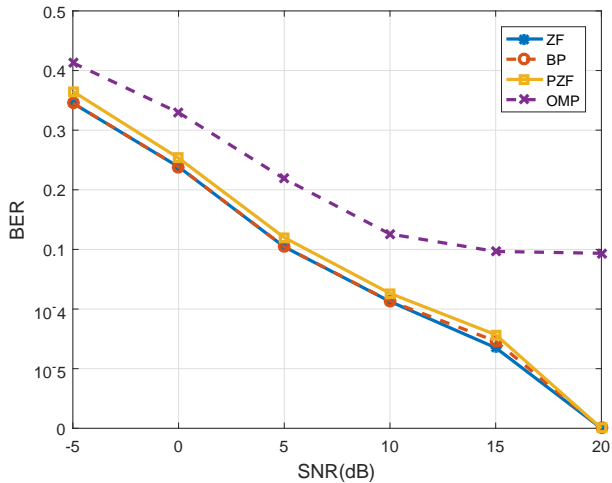}}
\caption{BER versus SNR with $K = 3$.}
\label{fig5}
\end{figure}

The relationship between the BER and the number of users is evaluated in Fig.~\ref{fig4}. The input data streams of the base station are QPSK signals, and the maximum likelihood estimation is adopted at the receiving end. The BER is averaged over   data streams transmission. As shown in Fig.~\ref{fig5}, the BER of the proposed algorithm has similar performance to ZF precoding, especially when the number of users is large, which is superior to the other two hybrid precoding schemes.

To learn the impact of the signal-to-noise ratio(SNR) on the BER thoroughly, we illustrate the BER with respect of SNR in Fig.5, where three users is added. In the case of high SNR, the BER of our proposed algorithm can be consistent with the ZF precoding. As the SNR decreases, the BER will increase slightly, but shows significant advantages compared with the OMP and PZF algorithms whose performance loss is severe. This indicates that the hybrid precoding algorithm based on complex BP neural network can adapt to worse communication scenarios.

\section{Conclusion}
This paper introduces a novel neural-network-based hybrid precoding algorithm for mm-Wave multiuser massive MIMO scenarios. The complex BP neural network is studied. We present the activation function of the complex neural network and the gradient derivation of the back propagation process. In particular, the proposed hybrid precoding algorithm supports online learning. Simulation results show that the proposed algorithm have a better performance on spectral efficiency and BER than the current hybrid precoding algorithm based on mathematical methods, and can approach the full digital precoding.

\section*{Acknowledgment}
The authors would like to acknowledge the support from the National Key Research and Development Program of China under Grant 2017YFE0121600.

\end{document}